# Molecular-level understanding of the WGS and reverse WGS reactions on Rh through hierarchical multiscale approach

*by*

*Matteo Maestri*[1,2*] *and Karsten Reuter*[2]

[1]Laboratory of Catalysis and Catalytic Processes -Dipartimento di Energia

Politecnico di Milano

Piazza Leonardo da Vinci 32, I-20133 Milano, Italy

[2]Department Chemie

Technische Universität München

Lichtenbergstrasse 4, D-85747 Garching, Germany

*E-mail: matteo.maestri@polimi.it

**Abstract**

Hierarchically combining semi-empirical methods and first-principles calculations we gain a novel and noteworthy picture of the molecular-level mechanisms that govern the water-gas-shift (WGS) and reverse water-gas-shift (r-WGS) reactions on Rh catalysts. Central to this picture is that the WGS and r-WGS follow two different dominant reaction mechanisms: WGS proceeds according to a carboxyl (COOH) mechanism, whereas r-WGS proceeds according to a redox ($CO_2 \rightarrow CO + O$) mechanism.

The obtained results furthermore underscore the danger of common first-principles analyses that focus on *a priori* selected dominant paths. Not restricted to such bias, our herein proposed hierarchical approach thus constitutes a promising avenue to properly transport and incorporate the *ab initio* predictive-quality to a new level of system complexity.

*Keywords:*

water-gas-shift reaction; reverse water-gas-shift reaction; reaction mechanism; multiscale modeling; dominant reaction mechanism; first-principles calculations

## Introduction

Reforming technologies on noble metals are at the heart of converting fuels and biofuels to syngas and hydrogen and play also a crucial step in the valorization of non-food interfering biomasses (Inderwildi et al., 2008; Vlachos et al., 2010). To advance and rationally design these technologies, the reaction mechanisms enabled by the catalysts have to be understood at the atomic scale (Nørskov et al., 2009). Recent experimental and theoretical works have already made first progress towards a corresponding identification of the dominant reaction pathways involved in these processes (Horn et al., 2006; Maestri et al., 2009). Nevertheless, a deep understanding of the catalytic mechanism that would enable an atomic-scale understanding of the complex experimental evidence has not yet been reached. Specifically, a significant open issue is the kinetic role of $CO_2$ in reforming processes that is ultimately related to the assessment of the kinetic relevance of the water gas shift (WGS) reaction and its reverse (r-WGS) (Donazzi et al., 2010). The centerpiece for this is the identification of the dominant reaction pathways of $CO_2$ production and consumption that actually take place at specific operating conditions. Here, the great challenge is associated to the fact that the *dominant reaction mechanism* is inherently a multi-scale property of the system. It is the result of the interplay among *all* the possible events at the micro-scale (i.e., the *full* reaction mechanism) at the conditions dictated by the physics of the meso- and the macro-scales (Stampfl et al., 2002). The first crucial part to tackle this with computational theory would be the accurate description of the elementary steps at the micro-scale. This can in principle be achieved through electronic structure theory calculations that explicitly treat the electronic degrees of freedom and the quantum-mechanical nature of the chemical bonds. Nevertheless, for complex processes and "real" catalysts (e.g., presence of surface defects), the number of *potential* reaction steps to be accounted for at the micro-scale can be excessive, rendering a fully first-principles based analysis prohibitive. To avoid this, the usual approach is to scrutinize different putative dominant reaction mechanisms and compare them in order to conclude which one will eventually occur (Gokhale et al., 2008). On one side, this limits the number of steps to be scrutinized. On the other side, the intrinsic drawback is that the analysis is strongly biased by the initial selection of possibly dominant mechanisms.

In this work we approach the problem of the identification of the dominant reaction mechanism from a different and more general perspective. According to a hierarchical approach (Maestri, 2012; Salccioli et al., 2011), we first employ the hybrid UBI-QEP semi-empirical method for the evaluation of the kinetic parameters of *all* the steps at the micro-scale at negligible computational costs, but at controlled uncertainty (Maestri and Reuter, 2011). This undemanding estimate of the activation energies serves as initial approximate input for the identification of the dominant mechanism through reaction path analysis (RPA). As a result, this avoids the need of making an *a priori* selection of the dominant paths. Then, following a hierarchical methodology, by comparison with selected experiments we perform a systematic first-principles investigation of only the key-issues identified by the semi-empirical analysis with explicit density-functional theory (DFT) calculations. As such, the latter are most effectively utilized, since there is detailed guidance as to what the most relevant issues are. In doing so, we gain a novel picture of the molecular-level mechanisms that govern WGS and r-WGS on Rh, which are for the first time able to rationalize the experimental evidence.

**Methods**

We focus our analysis on the WGS and r-WGS experiments in an annular reactor of (Donazzi et al., 2008; Donazzi et al., 2008). For the semi-empirical description of the microscale, we use the microkinetic model of (Maestri et al., 2009) within a 1D heterogeneous model of the annular reactor (Maestri et al., 2008). Then, at the specific operating conditions of the experiments, we identify the dominant reaction mechanism through reaction path analysis (RPA) based on the actual species consumption rate (Maestri et al., 2008). By considering the actual species consumption rate, we account both for the effect of rate coefficients (pre-exponential and activation energy) and for the species surface coverage. These dominant reaction pathways are then refined with explicit DFT calculations. The DFT calculations are carried out with the ultrasoft pseudopotential code CASTEP (Clark et al., 2005), using the semi-local PBE functional (Perdew et al., 1996) to treat electronic exchange and correlation, and supercell geometries with relaxed three layer slabs, 12Å vacuum separation and a (2x2) surface unit-cell periodicity. Systematic convergence tests showed that for these cells the use of (6x6x1) Monkhorst-Pack grids (Monkhorst and Pack, 1976) for the Brillouin zone integrations and a plane wave expansion up to 400 eV ensures numerical convergence of the computed binding energies within ±30 meV. The explicit DFT activation barriers are determined by first locating the TS with climbing-image nudged elastic band, followed by a refinement and vibrational validation within the conjugate-gradient quadratic synchronous transit method (Govind et al., 2003; Henkelman et al., 2000).

## Results

### *Semi-empirical microkinetic analysis*

We focus our analysis on the WGS and r-WGS experiments of (Donazzi et al., 2008; Donazzi et al., 2008). At conditions as in the experiments, we first identify the dominant reaction pathways using a semi-empirical microkinetic model for the description of the micro-scale (Maestri et al., 2009).

As described in (Maestri et al., 2009), this detailed microkinetic model includes several mechanisms for the WGS reaction such as redox, carboxyl (COOH) and formate (HCOO) mechanisms, as well as the direct oxidation of CO with OH species. Out of the comprehensive list of potential reaction steps at the micro-scale, RPA identified as the dominant pathways the ones reported in Figure 1 (a and b).

**[FIGURE 1]**

The semi-empirical analysis predicts that for the WGS system (panel a), $H_2O$, upon adsorption on the catalyst, dissociates to OH and H. Then, CO adsorbs on the surface and converts to $CO_2$ through direct oxidation with OH. Water dissociation turns out to be the rate-determining-step (RDS) of the reaction network, while all the other dominant steps turn out to be at partial equilibrium. For temperatures higher than 450°C, the surface turns out to be reasonably free (free sites > 70%).

In order to derive the observed reaction order we follow the same procedure adopted by (Maestri et al., 2008). For the enumeration of the elementary steps we refer to the original microkinetic model (Maestri et al., 2009).

Consequently:

$$r_{WGS,obs} = r_{RDS} = k_7 \theta_{H_2O} \theta_{Rh} \qquad (1)$$

where $\theta$ the coverage of the adsorbed species at the catalyst surface (mean-field approach).

Since all the other dominant steps of Figure 1 (panel a) are the partial equilibrium (i.e., $r_{forward} \approx r_{backward}$) then:

$$k_{13} p_{H_2O} \theta_{Rh} \approx k_{14} \theta_{H_2O} \qquad (2)$$

By combining (1) and (2) one can get:

$$r_{WGS,obs} = r_{RDS} \approx \frac{k_{13}}{k_{14}} p_{H_2O} (\theta_{Rh})^2 \qquad (3)$$

and, since the free sites are higher than 70%, Eq. 3 can be approximated wih:

$$r_{WGS,obs} = r_{RDS} \approx \frac{k_{13}}{k_{14}} p_{H_2O} (\theta_{Rh})^2 \approx K_{WGS} p_{H_2O} \qquad (4)$$

Eq. 4 reveals an overall first order dependence with respect to $H_2O$ and a kinetic independence from CO.

For the r-WGS system (panel b), RPA shows that the same dominant pathways as in the WGS are followed in the reverse way. In particular, $CO_2$ converts to CO via reaction with H and the RDS is water formation ($OH+H \rightarrow H_2O$). Following the analogous procedure as done for WGS (i.e., Eqs. (1-4)), this leads to a complex kinetic dependence of the overall rate expression on $H_2$, $CO_2$ and CO. This is at variance with experiment, which reports a first order dependence with respect to $H_2O$ and $CO_2$ for WGS and r-WGS, respectively (Donazzi et al., 2008). In particular, this implies that the RDS of the network cannot be water formation for r-WGS.

*DFT refinement of the dominant pathways*

In order to understand the reasons behind this disagreement, we proceed by scrutinizing the identified dominant pathways (Figure 1 a and b) through explicit DFT calculations. Intriguingly, for the reaction pathways that govern the CO oxidation via OH, this first-principles refinement reveals that, contrary to what was included in the semi-empirical microkinetic model, the direct oxidation of CO by OH ($CO+OH \rightarrow CO_2+H$) is not an elementary step. Thus, it cannot be considered as an alternative route to the carboxyl mechanism. In fact, by the identification of the minimum energy path (MEP) between reactants and products, we find that such pathway instead proceeds through the formation of different stable reaction intermediates. First, a cis-carboxyl (COOH) forms. Then, the cis-carboxyl isomerizes to trans-carboxyl, before decomposing to $CO_2$ and H. This peculiar insight is only achieved by virtue of the first-principles analysis, as the semi-empirical UBI-QEP method used for the initial screening is blind with respect to the true nature of the transition state and cannot differentiate whether a reaction step is elementary or not. In fact, corresponding carboxyl pathways were originally included in the microkinetic model, but were completely ruled out by the effective semi-empirical energetics of the *false elementary step*. In this respect, it is very important to note that the peculiar information from the first-principles analysis was not concerning the specific value of the energetics assigned to a specific reaction step (that would have been in any case affected by the "material gap" issue), but was related to the proper description of the elementary steps at the microscale. This example clearly shows that in microkinetic model development not only proper values for kinetic parameters count, but also the elementary steps themselves accounted for the description of the microscale play a crucial role in correctly describing the established dominant pathways and the observed kinetic behavior.

*DFT-refined microkinetic analysis*

Following the results from the first-principles analysis we remove this non-elementary step from the full microkinetic model. In order to understand the effect of this modification we repeat the RPA for both WGS and r-WGS. As a result of the improved description of the micro-scale, WGS and r-WGS now follow different *dominant reaction mechanisms* under the investigated experimental conditions. Specifically (Figure 1 c and d), in WGS water activation is the RDS and CO converts to $CO_2$ through a carboxyl pathway. Since, all the steps in Figure 1c except for the RDS turn out to be at partial equilibrium, still the observed reaction rate is given by Eq. 3, in agreement with observed reaction order.

During r-WGS, instead, the dominant reaction mechanism according to the DFT-refined microkinetic model (Figure 1 d) turns out to be different from the semi-empirical one (Figure 1 b). Specifically, $CO_2$ converts to CO via decomposition to CO and O, skipping the carboxyl-pathway. Moreover, $CO_2$ decomposition turns out to be the RDS. Consequently, the observed reaction rate is given by:

$$r_{r-WGS,obs} = r_{RDS} = k_{23}\theta_{CO_2}\theta_{Rh} \qquad (5)$$

Since the model predicts the $CO_2$ adsorption step to be at partial equilibrium at the investigated conditions, one finally ends up with:

$$r_{obs} = r_{RDS} \approx \frac{k_{21}}{k_{22}} p_{CO_2}(\theta_{Rh})^2 \approx K_{r-WGS}p_{CO_2} \qquad (6)$$

These results lead to a clear first-order kinetic dependence on $H_2O$ and $CO_2$ for WGS (Eq. 4) and r-WGS (Eq. 6), respectively, and thus reconcile this new picture of the dominant mechanisms at the molecular level with the experimentally observed macroscopic reaction orders for WGS and r-WGS.

## Conclusions and outlook

As a whole, efficiently exploiting the predictive power of first-principles methods within a hierarchical multiscale methodology, we derive a novel mechanistic picture of WGS and r-WGS reactions on Rh catalysts. Our findings point out that WGS and r-WGS proceed according to two different dominant reaction mechanisms. In particular, WGS follows a carboxyl mechanism in analogy with what was determined for Cu and Pt (Gokhale et al., 2008; Grabow et al., 2008). Instead, r-WGS follows in the reverse way a redox mechanism. This new mechanistic description naturally explains all experimentally observed reaction orders.

On a more general basis, the present work demonstrates:

1) the critical importance of an accurate and comprehensive description of the elementary steps to be included at the micro-scale for the molecular-level understanding of the dominant mechanisms that establish at the meso- and macro-scales
2) the crucial difference between *microkinetic model* and *dominant reaction mechanism*. The first one is property only of the microscale (making and breaking of chemical bonds at the surface) and has to account for all the potential *elementary* steps. The second-one is an inherent *multiscale* property of the reacting system and determines the actual catalytic cycle for specific operating conditions. The consistency between the properties of the predicted catalytic cycles and the observed macroscopic kinetic behavior is of utmost importance in developing predictive microkinetic models
3) for complex processes, an exhaustive, unbiased treatment of the microscale would require the evaluation of an unfeasible number of potential steps that can occur at the surface. In this respect, we herein provide a show-case of how first-principles calculations and semi-empirical methods can be hierarchically employed in conjunction with kinetically relevant experiments for the microkinetic analysis of complex chemical kinetics – without the need of making *a priori* assumptions on the dominant reaction mechanism at the atomic scale.

**Acknowledgements**

M.M. gratefully acknowledges a fellowship from the Alexander von Humboldt Foundation. Computational time at Leibniz Supercomputing Center (Munich, Germany) is gratefully acknowledged (project pr47ma).

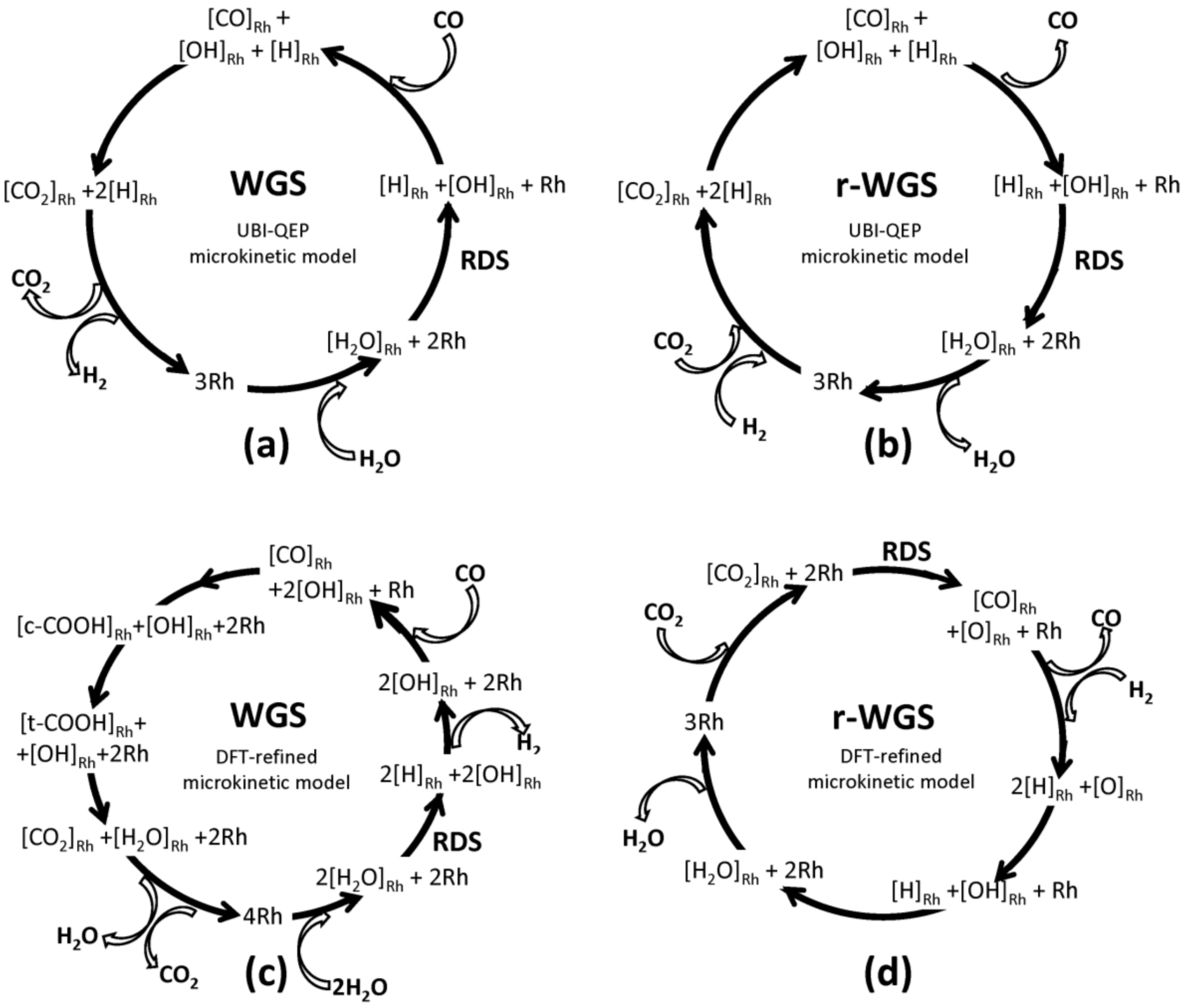

**Figure 1**

Dominant reaction mechanisms for WGS and r-WGS on Rh(111). Panels (a) and (b): UBI-QEP semi-empirical microkinetic model; panels (c) and (d): hierarchically DFT-refined microkinetic model. $[X]_{Rh}$ means adsorbed species.